\documentclass{svjour2arxiv}                    
\usepackage{graphicx}
\usepackage{hyperref}
\usepackage{cite}
\newcommand{\be}{\begin{equation}}
\newcommand{\ba}{\begin{eqnarray}}
\newcommand{\ee}{\end{equation}}
\newcommand{\ea}{\end{eqnarray}}
\begin{document}

\title{Investigating puzzling aspects of the quantum theory by means of
its hydrodynamic formulation}


\author{A. S. Sanz}


\institute{
           Instituto de F{\'\i}sica Fundamental (IFF-CSIC) \\
           Serrano 123, 28006 Madrid, SPAIN \\
           Tel.: +34-91-561-6800, ext.~941113\\
           Fax:  +34-91-585-4894 \\
           \email{asanz@iff.csic.es}}

\date{}

\maketitle

\begin{abstract}
Bohmian mechanics, a hydrodynamic formulation of the quantum theory,
constitutes a useful tool to understand the role of the phase as the
mechanism responsible for the dynamical evolution displayed by quantum
systems.
This role is analyzed and discussed here in the context of quantum
interference, considering to this end two well-known scenarios, namely
Young's two-slit experiment and Wheeler's delayed choice experiment.
A numerical implementation of the first scenario is used to show
how interference in a coherent superposition of two counter-propagating
wave packets can be seen and explained in terms of an effective model
consisting of a single wave packet scattered off an
attractive hard wall.
The outcomes from this model are then applied to the analysis of
Wheeler's delayed choice experiment, also recreated by means of a
reliable realistic simulation.
Both examples illustrate quite well how the Bohmian formulation helps
to explain in a natural way (and therefore to demystify) aspects of the
quantum theory typically regarded as paradoxical.
In other words, they show that a proper understanding of quantum phase
dynamics immediately removes any trace of unnecessary artificial
wave-particle arguments.

\keywords{Bohmian mechanics \and quantum phase \and velocity field
\and interference \and Young two-slit experiment \and Wheeler
delayed-choice experiment}

\end{abstract}


\section{Introduction}
\label{sec1}

Quantum phenomena occur in real time.
Although this may seem a trivial statement, monitoring the evolution in
time of quantum systems in the laboratory has not been a feasible task
until recently.
The development and improvement of highly refined experimental
techniques have allowed us to explore time domains of the order of the
femto- and attoseconds, at which the dynamics of many quantum processes
and phenomena of interest take place (e.g., change of molecular
configurations, charge transfer, entanglement dynamics, electron
ionization by very intense laser fields, diffusion of adsorbed
particles on surfaces, etc.).
However, to obtain a full picture of quantum systems, it is necessary
to perform a large number of measurements over hypothesized identical
realizations of the same experiment or, equivalently, over many
identical systems (of course, by ``identical'' it should be understood
``nearly identical'' in both cases).
Statistics is the only way to extract relevant (i.e., physically
meaningful) information from quantum systems, something that also
happens when dealing with classical ensembles.
It is a full collection of statistical data what constitutes the
outcome compatible with the solutions provided by Schr\"odinger's
equation.
Therefore, any discussion about how every single event from within
such a data collection evolves in time turns out to be nonphysical
and, perhaps, even meaningless. But, is this totally true?

Young's two-slit experiment constitutes an ideal candidate to tackle
the above question.
In a real laboratory performance of this experiment, a beam of
identical particles is launched against the two slits, observing far
away behind them the appearance of the well-known interference fringes
---an alternating pattern of regions of maximum and minimum density of
recorded events (detected particles).
According to Dirac \cite{dirac-bk}, each one of these particles
displays a wave-like nature, passing through {\it both} slits at the
same time and interfering with themselves ({\it self-interference})
behind them.
This model explains the observation of interference fringes.
In addition, from von Neumann's collapse hypothesis \cite{vonNeumann-bk}
it follows that, at the detector, the particle wave function collapses
at some random location.
That is, the particle exhibits its corpuscular nature and behaves as a
localized ``piece'' of matter.
Though odd, nowadays Dirac's reasoning (plus the collapse postulate)
constitutes the most widespread conception of how quantum systems
behave.
This oddity becomes even more striking after a closer look at the
experiment, where the solutions described by quantum mechanics are
built up particle by particle (event by event), keeping no coherence
in time between two consecutive particles, even though they all come
from the same source \cite{pozzi}.
That is, particles are totally uncorrelated and, therefore, it is
not possible to explain quantum interference by appealing to former
physical interactions among them (entanglement) at the source.
Experiments with photons \cite{parker,rueckner,weis}, electrons
\cite{merli,tonomura}, ultracold atoms \cite{shimizu}, or even with
large molecular systems \cite{arndt1,arndt2} have all shown the
universality of quantum interference as a phenomenon emergent from
statistics, regardless of the size and complexity of the system
investigated.

Such kind of experiments invites in a natural way to formulate and
investigate descriptions of quantum phenomena in terms of statistical
single-event realizations, implementing realistic numerical simulations
of the experiment in order to gain some insight on the physical
mechanics underlying the quantum phenomena investigated.
In this regard, Bohmian mechanics, a hydrodynamic formulation of
the quantum theory \cite{sanz-bk1,sanz-bk2}, constitutes a reliable
and useful tool, where the evolution of quantum systems is represented
in terms of streamlines.
From a dynamical viewpoint, this formulation gives more relevance to
the the quantum phase (and hence the quantum current density) than to
the probability density.
This pragmatic and natural use of Bohmian mechanics is analogous to the
use of characteristics in other fields of physics and chemistry as an
analytical tool \cite{sanz-2014}, having nothing to do with the common
view that Bohmian trajectories constitute some kind of ``hidden''
variables \cite{bohm1,bohm2,bohm3}.
Bearing this in mind, here I analyze and discuss the role of the
quantum phase as a mechanism involved in the dynamics displayed by
quantum systems in the context of interference phenomena.
Accordingly, it is observed that this phenomenon is analogous to
dealing with effective barriers in Young-type experiments, in
compliance with recently reported data on this experiment
\cite{steinberg}.
This study is subsequently used to analyze Wheeler's delayed choice
experiment \cite{wheeler-bk} in terms of a realistic numerical
simulation, which removes any trace of paradox and explains in
simple terms what happens inside the interferometer.

The remainder of this work has been organized as follows.
The essentials of Bohmian mechanics and its contextualization with
respect to the quantum theory are introduced and discussed in
Section~\ref{sec2}.
In Section~\ref{sec3}, the role of the quantum phase in relation to
interference phenomena is discussed, introducing a new physical
understanding of the notion of (quantum) superposition as well as
the concept of effective dynamical potential (not to be confused with
Bohm's usual quantum potential).
In Section~\ref{sec4}, a numerical simulation of Wheeler's delayed
choice is analyzed taking into account the discussion of the previous
section.
Finally, some concluding remarks are summarized in Section~\ref{sec5}.


\section{Bohmian mechanics}
\label{sec2}

Quantum mechanics admits different formulations.
Each one emphasizes a way to conceive the quantum system and its
evolution in time, although they all are equivalent ---something
similar can also be found in classical mechanics.
For instance, while Schr\"odinger's wave mechanics allows to visualize
the time-evolution of quantum systems, Heisenberg's matrix formulation
provides us a point of view closer to that of classical mechanics,
since the role of the classical variables is taken by the quantum
operators.
Dirac's formulation establishes a bridge between both and is of
interest when dealing with open quantum systems, although Feynman's
path representation is more powerful computationally.
To establish a direct connection between quantum and classical systems
(quantum-classical correspondence), we additionally have phase-space
representations, such as the Wigner-Moyal or the Husimi ones.

In the particular case of Bohmian mechanics, what we have is a
hydrodynamic description of quantum systems, where the system
probability density is understood as a kind of fluid that spreads
throughout the corresponding configuration space.
Accordingly, there is an associated advective flux, namely the
probability current density, which is a manifestation of a given
velocity field acting on the probability density.
This can easily be seen by substituting the wave function in polar
form,
\be
 \Psi({\bf r},t) = \rho^{1/2}({\bf r},t) e^{iS({\bf r},t)/\hbar} ,
\ee
into Schr\"odinger's equation (let us consider here only the
nonrelativistic scenario for simplicity),
\begin{equation}
 i\hbar\frac{\partial \Psi}{\partial t} =
 \left( -\frac{\hbar^2}{2m}\nabla^2 + V \right) \Psi .
\end{equation}
This gives rise to two coupled, real-valued differential equations.
One of them is the usual continuity equation,
\be
 \frac{\partial \rho}{\partial t} + \nabla {\bf J} = 0 ,
 \label{ce}
\ee
where ${\bf J} = \rho \nabla S/m$ is the probability current density
playing the role of the aforementioned advective flux, associated with
the velocity field ${\bf v} = \nabla S/m$.
The other equation can be considered to be a quantum version of the
Hamilton-Jacobi equation,
\be
 \frac{\partial S}{\partial t} + \frac{(\nabla S)^2}{2m}
  - \frac{\hbar^2}{2m} \frac{\nabla^2 \rho^{1/2}}{\rho^{1/2}} + V = 0 ,
 \label{qHJe}
\ee
where the third term is Bohm's quantum potential.
This equation led Bohm to postulate the existence of trajectories that
could be identified with the actual particle motion, constituting a set
of underlaying {\it hidden variables} that would then explain the causal
evolution of the quantum system, although they would not be accessible
to the experimenter.
Based on Eq.~(\ref{ce}), however, there is no necessity to establish
such identification; the existence of a current ${\bf J}$ by itself
allows us to define streamlines to analyze the evolution of the
quantum system, as we also do when dealing with classical fluids or,
in general, transport phenomena.
These streamlines or trajectories are obtained after integrating the
equation of motion
\begin{equation}
 {\bf v} = \dot{\bf r} = \frac{\nabla S}{m} = \frac{\bf J}{\rho}
   = \frac{\hbar}{2im}\ \! \nabla
     \ln \left( \frac{\Psi}{\Psi^*} \right) .
 \label{vel}
\end{equation}

The main goal of the Bohmian formulation consists in dealing with
quantum systems as if they were a kind of fluid given their
delocalization in the corresponding configuration space.
Swarms of streamlines or trajectories provide us with statistical
information on how such a fluid evolves, indicating which regions of
the configuration space are more highly populated or avoided at each
time (i.e., where the probability density is higher or lower,
respectively).
Rather than true paths pursued by the system, such trajectories should
be identified with the paths followed by some ideal tracer particles
that move with the associated quantum flow \cite{sanz-ajp2012}, thus
providing information about the latter ---in the same sense that a leaf
on a river tells us about the dynamics of the water flow, but does not
reveal any information about the motion of the individual water
molecules that constitute it.
Nonetheless, the strength of this representation relies on its
closeness to classical statistical treatments, where physically
meaningful quantities arise from ensembles rather than from single
trajectories.
Now, although physically irrelevant, such single trajectories are
useful to infer properties associated with the system or process under
study (in chemical reactivity, for instance, these trajectories allow
to ascertain whether certain initial conditions lead to the formation
of products or not).
This is precisely the kind of information that can be expected from
Bohmian trajectories, which is typically ``hidden'' within other
formulations of quantum mechanics, although not incompatible with
them at all.
Of course, this idea transcends Bohmian mechanics; in the literature,
it has been used with analogous purposes in different areas of physics
and chemistry \cite{sanz-2014}.
It is also in this sense that it would not be appropriate to consider
Bohmian trajectories as hidden variables, because we can find exactly
the same description in other fields.


\section{Quantum interference, phase dynamics and the Bohmian
non-crossing rule}
\label{sec3}

\begin{figure}[t]
 \begin{center}
 \includegraphics[width=1.00\textwidth]{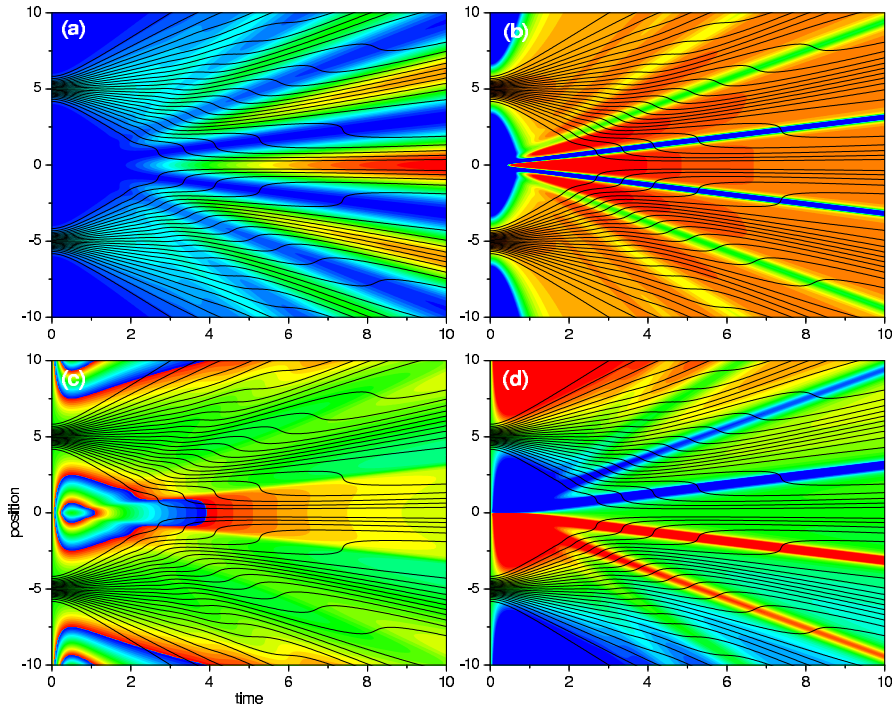}
 \caption{Contour-plots of the probability density (a), quantum
  potential (b), quantum phase (c), and velocity field associated (d)
  with a coherent superposition of two Gaussian wave packets simulating
  Young's two-slit experiment.
  Sets of Bohmian trajectories leaving each slit are superimposed to
  provide a more vivid insight of the flow dynamics.
  In this simulation, the initial width of the wave packets is
  $\sigma_0=0.5$ and their centroids are at $|x_0|=5$ (arbitrary units
  are considered without loss of generality, with $\hbar=m=1$).}
 \label{fig1}
 \end{center}
\end{figure}

Young's two-slit experiment is commonly explained appealing to the
superposition principle: the waves diffracted by each slit superimpose
and, depending on their phase at a given point, they may give rise to
intensity maxima (equal phase at that point) or minima (different
phase).
This phenomenon is illustrated in Fig.~\ref{fig1} by means of a
numerical simulation, where only the time-evolution along the
transversal coordinate (parallel to the plane where the slits are
allocated) has been considered.
Specifically, the initial wave function, accounting for the diffraction
at the slits, is described by a coherent superposition of two Gaussian
wave packets \cite{sanz-bk2},
\be
 \Psi(x,t) \propto e^{-(x-x_0)^2/4\sigma_0\tilde{\sigma}_t} +
  e^{-(x+x_0)^2/4\sigma_0\tilde{\sigma}_t} ,
 \label{superp}
\ee
where $\sigma_0$ is the initial width of the wave packets (of the order
of the slit width) and $\sigma_t = |\tilde{\sigma}_t|$ the width at a
time $t$, with $\tilde{\sigma}_t = \sigma_0[1 + i(\hbar t/2m\sigma_0)]$.
Because the expression (\ref{superp}) for the wave function is fully
analytical, the field quantities represented in the panels of
Fig.~\ref{fig1} can be readily determined by substituting into the
corresponding (analytical) expressions the values of $x$ and $t$ where
we want to evaluate them.
As for the Bohmian trajectories, they have been numerically computed by
means of a simple 4th-order Runge-Kutta scheme, which takes advantage
of the analyticity of (\ref{superp}) to evaluate the right-hand side
of the guidance equation (\ref{vel}) at each time step.

The development of interference fringes as a function of time is shown
in the contour-plot of the probability density associated with
(\ref{superp}), $\rho = |\Psi|^2$, displayed in Fig.~\ref{fig1}a.
The Bohmian trajectories superimposed in the figure (black solid lines)
provide an accurate description of how such a probability density
evolves from two localized regions to separate fringes that cover a
larger area.
Notice that the probability density is not simply an abstract concept,
but has a very precise physical meaning: it tells how many events are
registered within a certain region at a given time.
This is in compliance with the fact that, in real life, detectors have
a finite width and, therefore, at each position they collect (during a
fixed time) a number of events proportional to $\rho$.
Numerical simulations aimed at providing a realistic description of
diffraction by different types of systems \cite{sanz-2000,sanz-2002}
show that, effectively, histograms built up with ensembles of Bohmian
trajectories reproduce the theoretical predictions obtained from $\rho$.

Typically, the mechanical explanation for the particular evolution
displayed by the trajectories relies on Bohm's quantum potential (see
Fig.~\ref{fig1}b).
This potential is considered to be the mechanism leading the
trajectories to eventually distribute along a series of plateaus and,
therefore, to observe maxima (densely populated regions of nearly free
motion), and minima (void regions between adjacent plateaus, where
quantum forces are very intense) \cite{sanz-2002}.
To some extent, this information is redundant, since the trajectories
are streamlines connected to the current density, and hence their
topology, will always be in agreement with how the latter evolves
(i.e., in principle there is no need for appealing to a quantum
potential).
What is not that trivial here, however, is the physical role of the
quantum phase, $S$, and its implications.
As seen in Fig.~\ref{fig1}c, independently of the value of $\rho$,
$S$ is well defined everywhere since the very beginning.
The meaning of coherent superposition is linked to this fact: two waves
are coherent if there is a continuity of phase, which makes impossible
to consider both waves as independent entities.
From this point of view, the longstanding debate about the role of the
observer in Young's experiment is totally meaningless: the observer
changes completely the experiment, breaking down such continuity of
phase, and therefore making impossible the detection of eventual
interference features.

\begin{figure}[t]
 \begin{center}
 \includegraphics[width=1.00\textwidth]{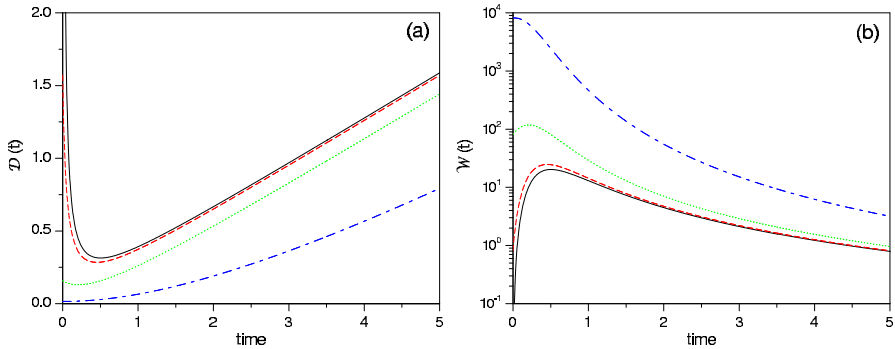}
 \caption{Time-dependence of the width (a) and well depth (b) of the
  effective interference dynamical potential associated with the
  coherent superposition of Gaussian wave packets displayed in
  Fig.~\ref{fig1} when both wave packets move initially towards each
  other.
  Each curve refers to a different value of the momentum associated
  with the centroids of the wave packets: solid black: $|p_0|=0$;
  dashed red: $|p_0|=1$; dotted green: $|p_0|=10$; dash-dotted blue:
  $|p_0|=100$.
  Other parameters are as in Fig.~\ref{fig1}.}
 \label{fig2}
 \end{center}
\end{figure}

Because of the non-additivity of $S$, there are two clearly
distinguishable dynamical regions, as seen in Fig.~\ref{fig1}d by
means of the associated velocity field, $v = \dot{x}$, specified
by Eq.~(\ref{vel}).
This naturally leads to conceive a single-slit model, where the flow
leaving one of the slits is reflected back by an effective potential
function that has nothing to do with the usual Bohm's potential.
Specific details of this model can be found in \cite{sanz-2008}.
Here it is enough to say that it consists of a square attractive well
followed by an impenetrable wall located at $x=0$.
The well depth ($\mathcal{D}$) and width ($\mathcal{W}$) not only
depend on time, but also on different initial physical parameters as
\be
 \mathcal{D}(t) = \frac{2\hbar^2}{m} \frac{1}{\mathcal{W}(t)} , \qquad
 \mathcal{W}(t) = \frac{\pi\sigma_t^2}{\displaystyle
  \frac{2|p_0| \sigma_0^2}{\hbar}
        + \frac{\hbar t}{2m\sigma_0^2}\ \! |x_0|} .
\ee
The second of these expressions has been obtained from a generalization
of the coherent superposition (\ref{superp}), where initially both wave
packets move towards $x=0$ at the same speed (the corresponding initial
momenta have the same absolute value, $|p_0|$, and opposite directions
\cite{sanz-2008}).
A plot of these two quantities for different values of $|p_0|$ is
displayed in Fig.~\ref{fig2}.
According to this simple scattering model, the initial coherence
induces an effect analogous to having two separate dynamical regions
that can be independently associated.
More importantly, the Bohmian trajectories coming from the upper slit
cannot cross the point $x=0$, and vice versa.
This allows us to state that, even if we know nothing about the true
individual motion of the quantum particles, at least at the level of
the wave function, fluxes do not mix.
This is precisely what Kocsis {\it et al.}~\cite{steinberg} observed
experimentally.
Although it is impossible to accurately determine the true path
pursued by a quantum particle, the fact that there is a continuity of
the average transverse momentum in space at a given time physically
means that quantum dynamics cannot be naively analyzed in terms of the
superposition principle.

The experimental results reported in \cite{steinberg} are in compliance
with the above model and the Bohmian formulation (although they are
far away from constituting a confirmation of the real existence of
Bohmian trajectories as the true paths followed by quantum particles).
Testing the direct equivalence between a two-slit experiment and the
scattering of a quantum system off an impenetrable attractive barrier
at a quantum level, as described above, perhaps involves a lot of
technical difficulties.
A feasible, worth pursuing substitute in this regard could be an
experiment in the line of those performed by Couder and Fort at the
CNRS (France) and Bush at the MIT (USA) with bouncing droplets
\cite{couder:Nature:2005,couder:PRL:2006,couder:JFluidMech:2006,couder:PNAS:2010,bush:PNAS:2010,bush:PRE:2013,bush:ARFM:2015}.


\section{Wheeler's delayed choice experiment revisited}
\label{sec4}

\begin{figure}[t]
 \begin{center}
 \includegraphics[width=1.00\textwidth]{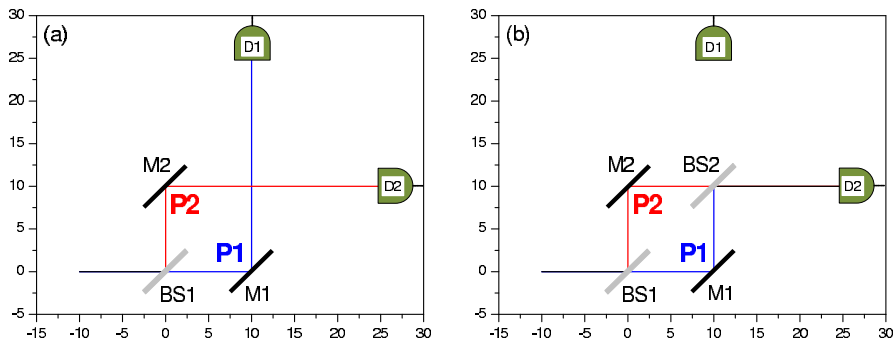}
 \caption{Traditional optical (geometric) ray representation of the
  two scenarios considered by Wheeler in his delayed choice experiment
  \cite{sanz-wheeler}.
  The photon can be either transmitted or reflected by the first
  beam splitter (BS1) with the same probability of 50\%.
  This generates two possible paths, P1 (blue) and P2 (red).
  (a) Open configuration: with absence of a second beam splitter, the
  probability to detect the photon at D1 or D2 is the same (50\%).
  (b) Closed configuration: if a second beam splitter BS2 is
  introduced, the photon behaves as a wave, which interferes
  constructively along P2 and destructively along P1.
  In this case, all the detections are registered at D2.}
 \label{fig3}
 \end{center}
\end{figure}

The previous results are very useful now to understand and explain in
a natural way Wheeler's delayed-choice experiment \cite{wheeler-bk},
removing any trace of paradoxical behavior.
With this thought-experiment Wheeler wanted to reformulate one of the
major issues of the Bohr-Einstein debates \cite{bohr-einstein}: when
does the quantum system make the choice to behave as a wave or as a
particle?
To this end, Wheeler conceived a clever experiment involving
an optical Mach-Zehnder interferometer with a movable second beam
splitter, and with a very dimmed light beam, so that at each time
there is one and only one photon passing through the device.
To understand the essence of the experiment and where the paradox
arises, let us focus on the traditional schematics of the two
interferometer configurations, displayed in Fig.~\ref{fig3}, where the
possible photon pathways are indicated in terms of optical (geometric)
rays (this is a typical representation).
Consider first that a photon enters the interferometer in the open
configuration, illustrated in Fig.~\ref{fig3}a.
The beam splitter BS1, oriented at 45$^\circ$ with respect to the
photon incidence direction, may produce either direct transmission
towards a mirror M1, along a path P1 (denoted by the blue line), or
a perpendicular deflection (reflection) towards a mirror M2, along P2
(red), with the same probability of 50\%.
In both cases, when the photon reaches the mirror (either M1 or M2), it
undergoes a deflection of 90$^\circ$ with respect to the corresponding
incidence direction.
Eventually, the photon arrival will be registered with the same
probability (50\%) either by a detector D1, along P1, or a detector
D2, along P2.
This is a typical scenario where the photon would exhibit its
corpuscular nature all the way through.
Next, the experiment is slightly modified, inserting a second beam
splitter, BS2, at the place where P1 and P2 intersect, as shown in
Fig.~\ref{fig3}b.
Moreover, to avoid any phase-difference, the path length along P1 and
P2 are the same between BS1 and BS2.
From a classical viewpoint, the lack of phase-difference produces that
all (classical) light would reach D2.
This result should be the same when the light beam is so weak that the
experiment is reproduced photon by photon, which means that photons
will be detected by D2.
To explain this result, it is necessary to assume that the photon
behaves as a wave.
Accordingly, the beam splitter BS2 separates the horizontal and
vertical wave components of the photon (just as BS1 did previously),
which may come from P1 or from P2.
After some simple algebra and the geometry of the setup, it is easy to
see that the vertical components associated with the paths P1 and P2
are out-of-phase ($180^\circ$) and cancel out, while the horizontal
components are in-phase and their addition results in constructive
interference, which explains why all photons are detected by D2.
These two scenarios allow us to observe the dual nature of photons
as well as, in general, any quantum particle.
The ``mystery'' posed by Wheeler comes when BS2 is introduced or
removed once the photon is already inside the interferometer.
Wheeler's answer to this situation is that, regardless of when BS2 is
put into play, the photon always behaves as it should, just like if it
could somehow anticipate what is going to happen in future (inserting
or removing BS2) and then behaving accordingly.
That is, the photon makes a ``delayed'' choice, ``taking its decision''
on which aspect it will display, corpuscle or wave, only when BS2
has been removed or inserted, respectively.
Nowadays this experiment is not a thought-experiment anymore; the
puzzling and challenging dual behavior of quantum particles has been
confirmed in the laboratory in many different ways
\cite{wheeler-exp1,wheeler-exp2,wheeler-exp3}.

\begin{figure}[t]
 \begin{center}
 {\includegraphics[width=6.75cm]{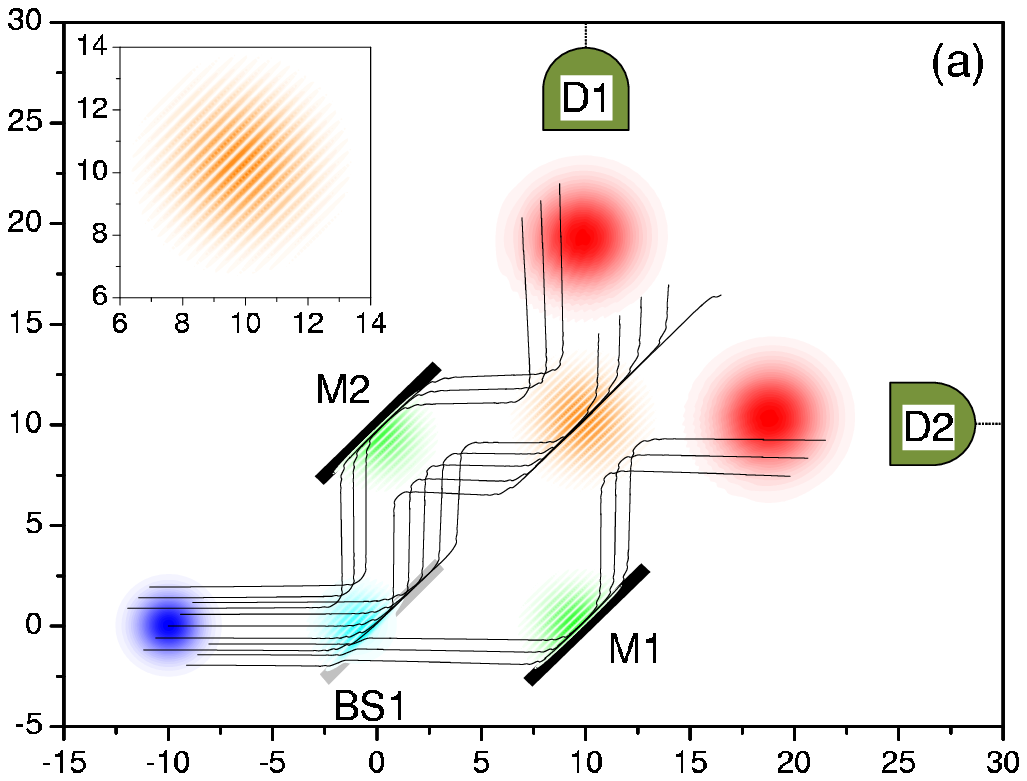}
  \vspace{10pt}}
  \includegraphics[width=6.75cm]{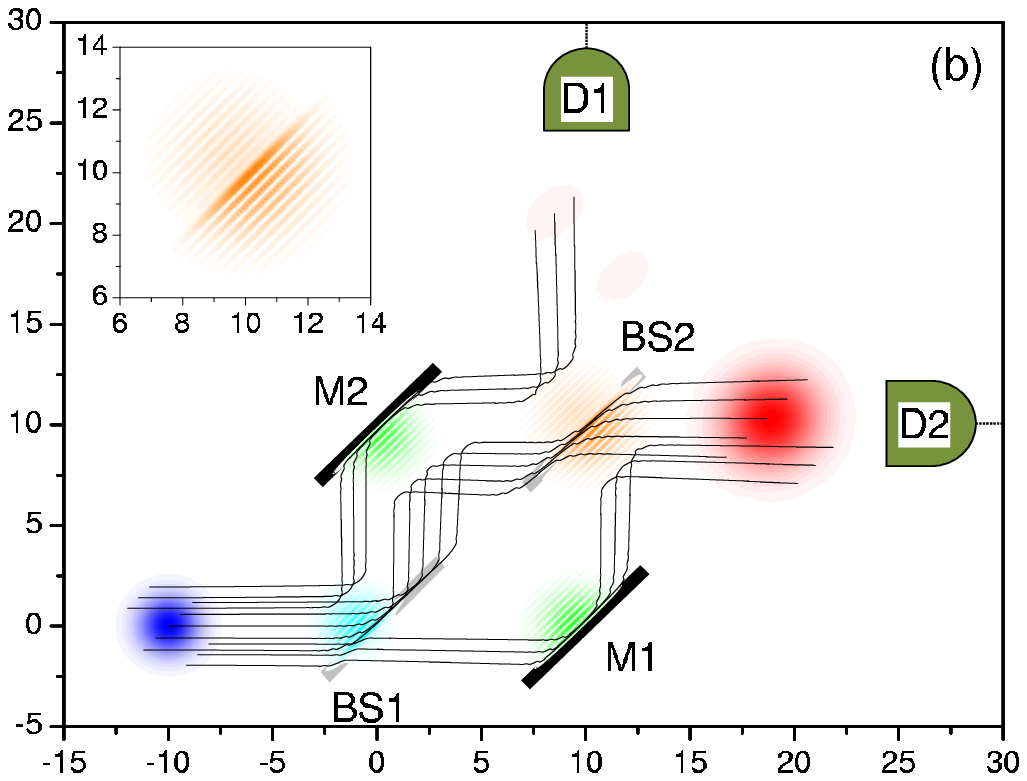}
 \caption{Numerical simulation of the two scenarios considered by
  Wheeler in his delayed choice experiment \cite{sanz-wheeler}:
  (a) open configuration and (b) closed configuration.
  The background monochrome contour-plots correspond to different
  stages of the wave-function evolution inside the interferometer:
  blue: initial state (Gaussian wave packet); light blue: splitting
  at BS1; green: reflection at the mirrors (M1 and M2); orange:
  superposition of the two wave packets at the position where BS2
  should be allocated; red: final stage (wave packets in their way
  to the corresponding detectors, D1 and D2).
  The insets show a magnification of the probability density in the
  region around BS2 in each case.
  The black solid lines denote ensembles of Bohmian trajectories
  starting with initial conditions covering different regions of the
  initial probability density (distribution).}
 \label{fig4}
 \end{center}
\end{figure}

The paradoxical behavior introduced by Wheeler readily dissipates
taking into account the phase dynamics discussed in the previous
section.
From a conceptual, Bohmian point of view, this experiment was firstly
discussed by Bohm and coworkers in 1985 \cite{bohm-1985}, and then
later on by Hiley and Callaghan \cite{hiley-2006}.
According to the Bohmian non-crossing rule \cite{sanz-2008}, appealed
by this authors, there is no paradox at all.
When BS2 is absent, because the trajectories coming from P1 and
P2 cannot cross the symmetry line at 45\%, those coming from P1 are
reflected in the direction of D2, and those from P2 in the direction
of D1.
That is, it is not that the photon follows P1 or P2 until it reaches
the corresponding detector, as it is usually argued to introduce the
corpuscular aspect, but there is an exchange in the directionality of
the associated quantum flows, typical of the collision of two coherent
wave packets \cite{sanz-2008}, as discussed in Section~\ref{sec3}.
On the other hand, when BS2 is introduced, even in the case that the
photon is already inside the interferometer, the recombination process
of the two waves that takes place around this beam splitter produces
that the two sets of trajectories will eventually go into only one of
the detectors.
This all-the-way wave behavior (notice that the traditional notion of
corpuscle just disappears) is illustrated in Fig.~\ref{fig4} by means
of realistic numerical simulations of the two processes described in
the preceding paragraph \cite{sanz-wheeler}.
Specifically, in this case, compared to the problem described in
Section~\ref{sec3}, the non-analyticity of the problem has led to
consider more robust calculations employing the split-operator
technique on a fixed grid to compute the evolution of the wave
function.
From this wave function, at each time step, the corresponding Bohmian
trajectories (denoted by black solid lines in both panels of
Fig.~\ref{fig4}) were synthesized on-the-fly by means of a 4th-order
Runge-Kutta method that was fed with interpolated values taken from
neighboring grid points (this method has been proven to be quite stable
and reliable in different atom-surface scattering problems
\cite{sanz-bk2}).
As it can be inferred from these simulations, the photon does not make
any choice at all.
What happens is that there is a modification of the boundary conditions
affecting its wave function, which simply gives rise to different
outcomes, regardless of whether BS2 is introduced or removed once
the wave function has started its evolution inside the interferometer.
This kind of realistic simulations are very important to better
understand the physics that is taking place in apparently paradoxical
situations, as it has also been recently shown, for example, in the
case of atomic Mach-Zehnder interferometry \cite{sanz-2015}, used to
discuss fundamental questions on complementarity
\cite{pritchard1,pritchard2}.

\section{Concluding remarks}
\label{sec5}

\begin{quote}
``[\ldots] we shall tackle immediately the basic element of the
mysterious behavior in its most strange form.
We choose to examine a phenomenon which is impossible, {\it absolutely}
impossible, to explain in any classical way, and which has in it the
heart of quantum mechanics.
In reality, it contains the {\it only} mystery.
We cannot make the mystery go away by ``explaining'' how it works.
We will just tell you how it works.
In telling you how it works we will have told you about the basic
peculiarities of all quantum mechanics.''
\end{quote}
These sentences start chapter~2 of the third volume of Feynman's
Lectures on Physics \cite{feynman-bk3}.
Effectively, the two-slit experiment probably constitutes the most
elegant manifestation of the quantum nature of material particles.
According to the traditional explanation of this experiment, what
happens is that the particle, at some point before reaching the slits,
behaves as a wave.
The two outgoing diffracted waves then recombine again, giving rise to
the typical interference fringes.
This notion of single-particle {\it self-interference} is what Feynman
had in mind when those above sentences were written, just the same as
many other of the founders of quantum mechanics before, starting from
Dirac, who stated that, in a beam of light consisting of a large number
of photons, each photon only interferes with itself and not with the
others \cite{dirac-bk}.
This notion has prevailed until today, but is there still room for
thinking quantum phenomena in a different way?

The different representations of the quantum theory provide us with
different aspects of this theory, something similar to what we already
know from the different classical approaches.
Bohmian mechanics constitutes one of these representations, which
stresses the role of the quantum phase, helping to understand how the
system evolves throughout the corresponding configuration space and
how the different elements (boundaries) influence its evolution.
In particular, we have focused on quantum interference, showing how
it emerges in Young's two-slit experiment and, based on the results
observed in this renowned experiment, we have also analyzed Wheeler's
delayed-choice experiment.
By analyzing the topology of the corresponding trajectories, it is
found that the phenomenon of quantum interference is analogous to
dealing with effective barriers, helping to provide mechanical
explanations and to remove paradoxical aspects of the quantum theory.
It is worth stressing that the same ``recipe'' can be (has been)
transferred to other fields of physics and chemistry with similar
purposes \cite{sanz-2014}, which leaves little room to keep thinking
Bohmian mechanics as a hidden-variable theory.


\begin{acknowledgements}

The author acknowledges support from the Ministerio de Econom{\'\i}a y
Competitividad (Spain) under Project No. FIS2011-29596-C02-01 as well
as a ``Ram\'on y Cajal'' Research Fellowship with Ref.~RYC-2010-05768.

\end{acknowledgements}




\begin{thebibliography}{}
\bibitem{sanz-wheeler}
 A more detailed work on Wheeler's experiment, including a removable
 beam splitter, is currently being prepared.
\end{thebibliography}
\end{document}